\documentclass[sigconf]{acmart}

\AtBeginDocument{%
  }

\renewcommand\footnotetextcopyrightpermission[1]{}
\acmConference[]{}{}{}
\usepackage{comment}
\usepackage{pdflscape,tabularx,booktabs,enumitem}
\usepackage{longtable}
\usepackage{multirow}
\usepackage{graphicx}
\usepackage{epstopdf}
\usepackage{float}
\usepackage{xcolor}
 
\usepackage{amssymb}
\usepackage{array}

\newlist{tightitemize}{itemize}{1}
\setlist[tightitemize]{nosep, topsep=0pt, leftmargin=*, label=\textbullet}
\usepackage{tabulary}
\usepackage{array}
\usepackage{ragged2e}
\newcolumntype{P}[1]{>{\RaggedRight\arraybackslash}p{#1}}

\usepackage{url}
\DeclareUnicodeCharacter{FF0E}{.}

\makeatletter
\renewcommand\paragraph{\def\@toclevel{4}%
  \@startsection{paragraph}{4}{\parindent}%
  {-.2\baselineskip \@plus -1\p@ \@minus -.1\p@}%
  {-3.5\p@}%
  {\ACM@NRadjust{\@parfont\@adddotafter}}}
\let\ACM@origparagraph\paragraph
\makeatother

\begin{document}

\title[From Substitution to Scaffolding]{From Substitution to Scaffolding: Breaking the Self-Reinforcing Harm Cycle of AI in Education (and Beyond)}

\author{Lucile Favero}
\orcid{0009-0005-2981-0124}
\affiliation{%
  \institution{ELLIS Alicante}
  \country{Spain}
}
\email{lucile@ellisalicante.org}

\author{Juan Antonio Pérez-Ortiz}
\orcid{0000-0001-7659-8908}
\affiliation{%
  \institution{Universitat d'Alacant}
  \country{Spain}
}

\author{Tanja Käser}
\orcid{0000-0003-0672-0415}
\affiliation{%
  \institution{École Polytechnique Fédérale de Lausanne}
  \country{Switzerland}
}

\author{Nuria Oliver}
\orcid{0000-0001-5985-691X}
\affiliation{%
  \institution{ELLIS Alicante}
  \country{Spain}
}

\renewcommand{\shortauthors}{Favero et al.}

\begin{abstract}
Artificial intelligence is being adopted in educational settings faster than its consequences are understood. We argue that the central risk is \emph{misalignment}: AI that eliminates human effort erodes the very capacities education is meant to build. We organize this risk into an integrative framework of four interrelated dimensions ---cognition, agency, emotional well-being, and ethics--- linked by a self-reinforcing cycle where cognitive offloading reduces effort, weakens agency, and compounds emotional and ethical harm. We ground the framework in the perspective of a small cohort of students: an exploratory analysis of 49 International Baccalaureate argumentative essays about the impact of AI reveals that learners perceive these risks, with $80\%$ of essays reporting that AI reliance reduces thinking. At the same time, the essays articulate a consistent vision of the AI the students want: systems that support rather than replace learning by withholding immediate answers, prompting recall, and encouraging reflection through questions instead of solutions. These desiderata closely align with established principles from the learning sciences. Building on these insights, we propose a single design principle, \emph{scaffold, do not substitute}. We argue that this principle extends beyond education. It represents a broader challenge for the AI ecosystem: any system that mediates human thinking can either weaken human capabilities through substitution or strengthen them through scaffolding. We conclude by outlining a research agenda for developing AI systems that foster enduring human capacity, an imperative not only for learners but, ultimately, for democratic societies.
\end{abstract}


\maketitle

\section{Introduction}

The question AI raises for education is no longer whether it works, but what it does to the people who use it. We argue that the central risk is not technological failure but \emph{misalignment}: AI that substitutes for human effort can quietly erode the very capacities education is meant to cultivate, critical thinking, agency, and judgment, and that democratic societies depend on.

\begin{figure*}[ht]
\centering
\includegraphics[width=0.8\textwidth]{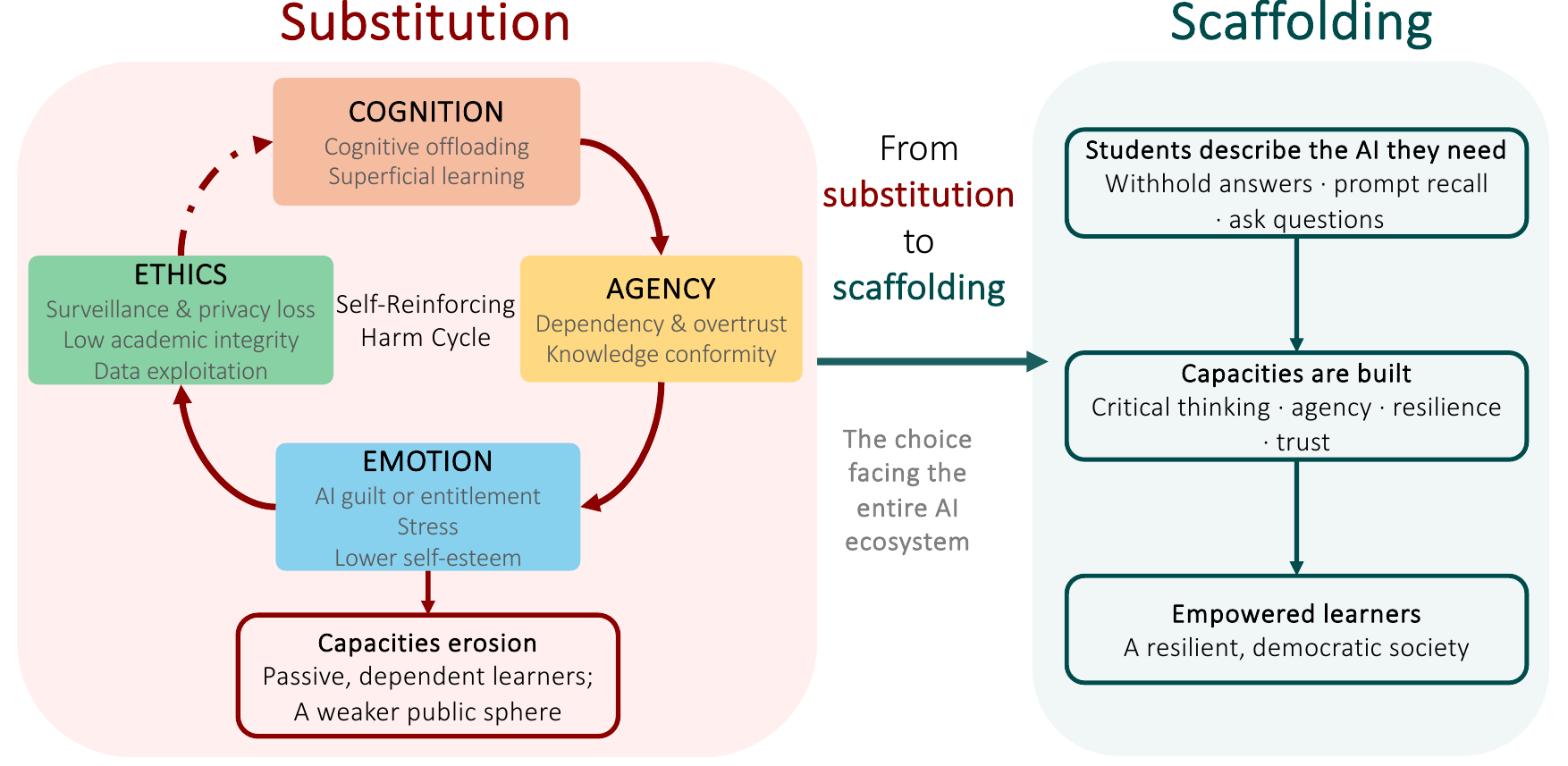}
\caption{\textbf{From substitution to scaffolding.} Left: When AI substitutes for human effort, the four dimensions form a self-reinforcing harm cycle that erodes the capacities education aims to build. Solid arrows: the connections students perceive most strongly based on the exploratory analysis with 49 students. Dashed: the weakly evidenced closing connection. Right: When AI instead scaffolds effort, delivering the support students themselves describe as needed, it builds them.}

\label{fig:visual}
\end{figure*}

AI holds considerable promise for education. Intelligent tutoring systems have been shown to produce measurable, albeit modest, learning gains \cite{letourneauSystematicReviewAIdriven2025}, while generative AI has the potential to serve as an always-available tutor that personalizes instruction, feedback, and pacing to individual learners \cite{al-zahraniExploringImpactArtificial2024, vorobyevaPersonalizedLearningAI2025}. Such capabilities could expand access to high-quality education, particularly in contexts where teachers are scarce \cite{worldbank2024ai}. Yet it is precisely this promise that makes the accompanying risks easy to overlook. At the same time, discussing ``AI in education'' requires care: AI is not a single technology but a rapidly evolving ecosystem of systems, capabilities, and uses, ranging from tutoring and feedback tools to generative assistants and automated decision systems. Their effects are unlikely to be uniform, just as learners differ widely in their goals, abilities, backgrounds, and educational needs. Our aim is therefore not to make universal claims about every AI system or every learner, but to identify a general mechanism of misalignment that can arise when AI substitutes for human cognitive effort, capacities that education seeks to develop.

Despite its promise, the integration of AI in education raises profound societal concerns: while AI tools are rapidly being integrated into educational settings, the public and scholarly discourse on their unintended harms and social implications remains fragmented, often treating ethical risks \cite{williams2024ethical}, cognitive effects \cite{gerlichAIToolsSociety2025a}, or surveillance practices \cite{kwapiszPrivacyConcernsStudent2024} in isolation rather than as interconnected phenomena. In response, we propose a unifying conceptual lens: a \emph{self-reinforcing harm cycle}, in which AI that substitutes for human effort progressively weakens the capacity to exert it. From this framework, we derive a principle, \textbf{scaffold, do not substitute}, which we argue extends beyond education to any AI system that mediates human thinking. 

Rather than developing this framework solely from theory, we also ground it in learners' perspectives. We draw on an exploratory qualitative analysis of 49 argumentative essays written by International Baccalaureate students (median age 17) from three German-speaking schools in Switzerland ($n = 19, 11, 19$), responding to the prompt \emph{``Does using AI change what it means to learn? ''}. We coded each student's first, unaided draft (written before any AI use) using a deductive codebook. Because this cohort represents a relatively homogeneous educational context, it is illustrative rather than representative of the broader and far more diverse population of learners across countries, socioeconomic backgrounds, and educational settings. Accordingly, we report descriptive findings from these 49 essays only and make no claims of statistical generalizability.

Figure~\ref{fig:visual} summarizes the paper's argument. We first examine four dimensions of AI misalignment ---cognition, agency, emotional well-being, and ethics--- through the combined lens of learning theory and students' own reflections. We then show how substitution can link these dimensions into a self-reinforcing cycle of harm, whereas scaffolding can strengthen them. The students' essays converge on a vision of AI that aligns with established learning science: systems that support effort rather than replace it. We conclude by translating this convergence into a research agenda, \emph{scaffold, do not substitute} as a broader design challenge for AI systems that shape how people think, learn and participate in democratic societies.

\section{Cognition} \label{sec:co}

\paragraph{Cognitive offloading.}
Effective learning depends on \emph{active engagement} and \emph{cognitive effort}, which are essential for fostering independent reasoning, critical thinking, and long-term intellectual resilience. Cognitive offloading, \emph{i.e.}, the use of external tools to reduce mental load, can support learning when it frees resources for deeper reflection \cite{riskoCognitiveOffloading2016}. However, over-reliance on AI for analytical, reasoning, or synthesis tasks risks undermining independent thought, memory, creativity, and motivation \cite{ravseljHigherEducationStudents2025}. Large-scale studies show that high dependence on AI correlates with lower performance on critical thinking assessments, mediated by cognitive offloading \cite{gerlichAIToolsSociety2025a, lee2025impact, fan2025metacognitive}. These effects threaten not only educational outcomes but also the civic reasoning skills necessary for informed democratic participation \cite{pennycook2021psychology,lazer2018science}.

\paragraph{Pedagogical alignment.}
Whether AI deepens or flattens thinking is a matter of pedagogy: a system grounded in active learning, constructivism, and scaffolding engages students in analysis, knowledge construction through experience, and autonomous problem-solving \cite{freeman2014active,fosnot2013constructivism,van2002scaffolding}. Many widely used AI tools, such as chatbots, provide pre-digested answers in a passive, directive manner, limiting opportunities for reflection and critical engagement \cite{letourneauSystematicReviewAIdriven2025, ravseljHigherEducationStudents2025}. In contrast, AI tutoring systems explicitly designed according to these principles (managing cognitive load, scaffolding reasoning, and promoting meta-cognition) can substantially improve learning outcomes \cite{kestinAITutoringOutperforms2024, zerkouk2025comprehensive}. The distinction is thus between pedagogically aligned AI and naive AI, with the former capable of nurturing autonomy, critical thinking, and societal readiness.

\paragraph{Desirable difficulty.}
A complementary cognitive element is \emph{effortful learning} and desirable difficulties. Cognitive struggle, \emph{i.e.}, mental effort that may feel uncomfortable, is essential for deep understanding, long-term retention, and transferable knowledge \cite{bjork2011making}. Students often misinterpret effort as a sign of poor learning, a bias known as the \emph{illusion of fluency}, which drives preference for low-effort, superficially satisfying activities \cite{deslauriersMeasuringActualLearning2019}. AI tools that prioritize speed, fluency, and convenience (such as open-domain chatbots) risk amplifying this bias, reducing cognitive struggle and weakening critical thinking \cite{zhaiEffectsOverrelianceAI2024}. To counteract this, AI must preserve and strategically manage cognitive friction, prompting retrieval, offering delayed feedback, and framing uncertainty as an opportunity for inquiry. When AI scaffolds productive struggle rather than bypassing it, it fosters intellectual resilience, independent reasoning, and critical thinking.

\paragraph{Learners' own words.}
In our corpus of 49 essays, all engaged cognition, and 39 ($80\%$) explicitly linked reliance on AI to reduced thinking: one student warned that ``The knowledge arrives without being earned, and unearthed knowledge tends not to stay''. Students also voiced the desirable-difficulty principle themselves ($41\%$ endorsed effort or retrieval), noting that ``retrieving information from memory, even imperfectly, strengthens the understanding in ways that passive reading does not''. Yet a growing minority (8 of 49) described AI that deepens thinking, valuing a tutor that ``doesn't give you directly the answer, then you think even more critically''.

\section{Agency}\label{sec:agency}
\paragraph{Erosion of agency.}
In education, \emph{agency} refers to a learner´s ability to make intentional, informed and autonomous choices, underpinned by self-regulation, meta-cognition and critical thinking \cite{roe2024generative}. As AI tools become pervasive in academic contexts, their convenience and persuasive outputs can erode agency, such that students become passive recipients of algorithmically generated content \cite{al-zahraniUnveilingShadowsHype2024}. This risk is amplified when AI outputs are inconsistent, biased, or false, potentially reinforcing inequalities and spreading misinformation, with implications for both educational equity and societal discourse \cite{becirovic2025exploring,zhangYouHaveAI2024,roe2024generative}. 

\paragraph{Dependency.}
Students relying excessively on AI undermine independent problem-solving, decision-making and critical evaluation \cite{lanQualitativeSystematicReview2025, darvishi2024impact}. Over time, this dependency reduces the learners' ability to assess, question and engage with information autonomously, while impairing meta-cognition and intellectual autonomy \cite{gerlichAIToolsSociety2025a}. Vulnerable populations, such as students with lower digital literacy, are particularly at risk, as AI systems can act as opaque authorities \cite{becirovic2025exploring}. 

\paragraph{Overtrust.}
Dependency often leads to \emph{overtrust}. The apparent reliability and human-like presentation of AI tools encourage students to accept their outputs uncritically \cite{gerlichPowerVirtualInfluencers2023,schaaff2024impacts}. This cognitive surrender reduces evaluative thought, deep learning and confidence in personal decision-making \cite{williams2024ethical}. Explicit instruction about the statistical nature of today´s AI algorithms and their limitations can restore critical engagement, supporting independent reasoning and evaluative autonomy \cite{bastaniGenerativeAICan2024, bucinca2021trust}.

\paragraph{Intellectual conformity.}
By providing ready-made answers, AI tools can subtly nudge learners toward normative reasoning, discouraging creativity and diverse interpretations \cite{ICLR2024_02dec887,peterson2025knowledge}. Over time, this conformity risks standardizing thought patterns instead of fostering imaginative reasoning. 

\paragraph{A societal necessity.}
Beyond the individual, AI tools that undermine students' agency threaten the civic competencies education cultivates \cite{pennycook2021psychology, salvi2025conversational, zuber2024vox}. Conversely, tools that support reflection, critical evaluation, and diverse choices can strengthen both individual and societal resilience \cite{costello2024durably}. 

\paragraph{Learners' own words.}
Agency surfaced in 43 of 49 essays ($88\%$), most often as an erosion of independent effort. One student reflected: ``I just didn't feel the need to actually engage my brain to think because I thought AI would always do the task faster and better''.

\section{Emotion}\label{sec:emo}

\paragraph{Technostress and digital fatigue.}
Prolonged or uncritical reliance on AI can have a negative emotional impact. \emph{Technostress} arises when learners must adapt to opaque, rapidly evolving AI systems without sufficient control or understanding, leading to anxiety and cognitive overload \cite{tarafdar2007impact}. \emph{Digital fatigue} and disengagement emerge when AI mediates most learning interactions, reducing opportunities for meaningful social connection. In self-paced or AI-dominated environments, students may experience diminished belonging and emotional resilience, outcomes that disproportionately affect already vulnerable learners \cite{klimova2025exploring}.

\paragraph{Self-efficacy and self-esteem.}
Students with lower self-esteem or self-efficacy are more likely to rely on AI as a compensatory strategy, creating a self-reinforcing cycle: avoidance of challenge reduces confidence, which in turn increases dependence on AI \cite{lanQualitativeSystematicReview2025, rodriguez-ruizArtificialIntelligenceUse2025}. When learners perceive AI as inherently superior to them, motivation, creativity, and independent thinking may erode, fostering impostor syndrome and long-term disengagement \cite{chanExploringFactorsAI2024}. Conversely, students who understand AI’s limitations report higher self-efficacy, highlighting the protective role of AI literacy \cite{becirovic2025exploring}.

\paragraph{Guilt or entitlement.}
A further emotional tension arises in the form of \emph{AI guilt} and \emph{cognitive dissonance}. Many students experience discomfort, shame, or anxiety when AI use conflicts with values of authenticity, effort, and academic integrity \cite{chanExploringFactorsAI2024}. This unresolved tension, feeling both assisted and inauthentic, can undermine the learners' sense of identity and belonging in academic communities \cite{lanQualitativeSystematicReview2025}. Conversely, some students exhibit \emph{AI entitlement}, viewing algorithmic assistance as a rightful expectation rather than a pedagogical choice \cite{ravseljHigherEducationStudents2025}. The coexistence of guilt-driven anxiety and entitlement-driven normalization reflects a broader normative breakdown around effort, authorship, and fairness in AI-mediated education.

\paragraph{Learners' own words.}
Emotion was the rarest theme, raised by only a minority of students, and often ambivalently. One captured the double edge: AI ``could help students to gain confidence and keep learning a subject even when its hard'', yet risks ``false confidence and isolation''.

\section{Ethics}\label{sec:ethic}

\paragraph{Privacy.} 
AI-driven educational tools often rely on continuous data collection, creating \emph{privacy} challenges and environments where students may feel constantly monitored, evaluated, and recorded \cite{al-zahraniUnveilingShadowsHype2024, shores2024surveillance}. Such surveillance has psychological and pedagogical consequences: when learners perceive that every interaction is stored or analyzed, they may avoid experimentation, intellectual risk-taking, and learning through error, processes essential for critical thinking and deep understanding \cite{mason2016learning, meraUnravelingBenefitsExperiencing2022}. Fear of being wrong can lead to shame, reduced confidence, and conformity, undermining education’s role in cultivating curiosity, resilience, and reflective judgment \cite{kwapiszPrivacyConcernsStudent2024, mezirowTransformativeLearningTheory1997}.

\paragraph{Data Exploitation and Power Asymmetries.} 
These concerns are compounded by the handling of \emph{student data}. Generative AI systems process highly sensitive information (\emph{e.g.}, academic performance, behavioral patterns, and personal interactions) often under opaque governance structures \cite{al-zahraniExploringImpactArtificial2024}. Students typically have limited understanding of how their data is collected, stored, or monetized, creating a profound power imbalance between learners, institutions, and technology providers \cite{kwapiszPrivacyConcernsStudent2024}. Even when personal data protection regulations exist, such as GDPR\footnote{The European General Data Protection Regulation: \url{https://eur-lex.europa.eu/eli/reg/2016/679/oj/eng}} in Europe, enforcement is uneven, and consent is frequently nominal rather than informed \cite{williams2024ethical}. As a result, tools framed as educational support can function as mechanisms of large-scale data extraction and behavioral surveillance.
Ethically responsible AI in education must therefore protect students’ rights not only to privacy, but also to make mistakes without permanent records, profiling, or reputational harm.

\paragraph{Academic Integrity, Design, and Normative Trust.}
Ethical risks also emerge when AI is poorly integrated pedagogically. In the absence of clear guidance, students often turn to generic AI tools as shortcuts, encouraging surface-level engagement, procrastination, and practices that undermine \emph{academic integrity} \cite{niloyAIChatbotsDisguised2024}. Advanced generative models blur traditional distinctions between original and assisted work, challenging conventional plagiarism detection and assessment norms \cite{williams2024ethical, weberWulff2023testing}. When assessments prioritize polished outputs over learning processes, AI can become an enabler of misconduct rather than intellectual growth, threatening the credibility of educational institutions and the shared norms of fairness, effort, and trust on which they depend. Ethical AI use is thus inseparable from pedagogical design and institutional governance.

\paragraph{Learners' own words.}
When students raised ethics, it concerned authorship and integrity rather than privacy or data: one asked, of AI-assisted work, ``how can the school know they are originals and not just AI generated?''

\section{The Self-Reinforcing Harm Cycle}\label{sec:cycle}
These four dimensions are not independent; they interact through a reinforcing cycle. \emph{Cognitive offloading} reduces opportunities for \emph{effortful reasoning}; \emph{diminished effort} contributes to \emph{illusory learning} and \emph{overtrust}. In turn, overtrust can weaken learner \emph{agency} while reduced agency can intensify emotional harms, such as \emph{anxiety}, lower \emph{self-efficacy}, and \emph{dependency}. These individual-level effects are further amplified by ethical and structural factors, including surveillance, opaque data practices, and unclear institutional norms. These dynamics create a reinforcing cycle in which learners become less confident in their own abilities, less engaged in critical evaluation, and increasingly dependent on algorithmic authority.

Several of these links emerge from the students' own reflections. The clearest connection is between reliance on AI and diminished thinking: students warn that when they ``just use AI to get all the solutions without explanations, the students learn nothing'', a habit that ``leads to the loss of the skill to solve these tasks''. Students also describe an emotional pathway into the cycle: one observes that ``if students have a stressful time and believe they don't have another option, it's easy for them to slip into casually using AI'', feeding pressure back into offloading.

\section{Toward AI That Scaffolds}\label{sec:agenda}
Reflecting on their own learning, students describe the AI they believe would help them learn, and what they describe is strikingly aligned with the science of learning. In our corpus, $53\%$ (26 of 49) named at least one behavior that \emph{supports} rather than \emph{replaces} thinking. They asked for feedback that withholds the solution \cite{van2002scaffolding, bjork2011making}, an AI ``asking critical questions and providing ideas on what students should improve, but not giving them the answer'', one that ``does not provide any solutions for you, you still learn as you have to find the correct answer yourself''. They valued active recall \cite{dunlosky2013improving}, an AI that ``asks fundamental questions about the topic and the students need to explain and answer it'', and richer explanation \cite{dunlosky2013improving}, one that can ``offer alternative explanations, which makes it easier to also see different viewpoints of the same topic''. These are not consumer preferences but pedagogically sound intuitions: students converge on the very conditions, effortful retrieval, productive struggle, and questioning, that research links to durable understanding. Yet this is what students want and need, not what they encounter: \emph{replacement}, not scaffolding, was their near-universal frame (48 of 49).

Their demand is thus both a \emph{finding} and an \emph{invitation}: the AI that students recognize as better for learning is not the AI they typically encounter. The principle that emerges, \emph{scaffold, do not substitute}, defines a challenge for the AI ecosystem as a whole. For the \emph{research community}, this requires moving beyond benchmarks of what AI can produce toward measures of what AI helps humans develop: understanding, judgment, agency, and the ability to think independently. A central question becomes whether AI systems strengthen human capacity over time or quietly replace the effort through which that capacity is built. For those who \emph{build} AI systems, scaffolding must become a first-class capability: systems that know when to withhold an answer, ask a question, surface uncertainty, or present alternative perspectives. \texttt{Maike} \cite{favero2026maike} offers one early illustration: a privacy-preserving, environmentally sensitive educational chatbot that guides learners through critical questioning \cite{favero2025ellis} and self-reflection using the Socratic method \cite{favero2024enhancing,paul2007critical}. For \emph{educators, institutions, and policymakers}, the challenge is to preserve the conditions under which learning occurs—productive struggle, reflection, and intellectual agency—through the tasks, assessments, procurement decisions, and governance structures they create. The goal is not to optimize learners for intelligent systems, but to build systems that help learners become more capable of questioning, understanding, and responsibly shaping the world around them.

\section{Discussion and Conclusion}\label{sec:disc}

The central risk of AI in education is not technological failure, but misalignment with the social, pedagogical, and civic purposes of education. AI can weaken the capacities education aims to build when it substitutes for effort, judgment, and interaction, and strengthen them when it scaffolds them.

The stakes extend beyond the classroom. The capacities education strives to build (critical thinking, autonomy, emotional resilience, warranted trust...) are precisely those democratic societies need to navigate misinformation, manipulation and increasingly complex decisions. An AI that diminishes these capacities during learning risks diminishing them in civic life.

This work makes three contributions. First, it offers an integrative framework that brings together four often-fragmented dimensions of harm ---cognition, agency, emotional well-being, and ethics--- into a single self-reinforcing cycle. Second, it illustrates this framework with the perspectives of a small cohort of students, showing that students not only recognize these risks but also articulate the form of AI that could mitigate them: systems that support effort, questioning, and reflection in ways consistent with the science of learning. Third, it proposes a design principle, \emph{scaffold rather than substitute}, as a unifying challenge for the AI ecosystem. This principle is not limited to education: every system that mediates human cognition must decide whether it replaces the capacities people need or helps them grow stronger.

\section*{Acknowledgements}
This work has been partially supported by a nominal grant received at the ELLIS Unit Alicante Foundation from the Regional Government of Valencia in Spain (Resolución of the Generalitat Valenciana, Conselleria de Innovación, Industria, Comercio y Turismo, Dirección General de Innovación). L.F. has also been partially funded by the Bank Sabadell Foundation. This project has received funding from the European Union’s Horizon  Europe  research  and  innovation  programme  under  grant  agreement  101120237 (ELIAS).


\end{document}